\pdfoutput=1
\documentclass[11pt,twoside,a4paper,cmspaper,final,collab]{cms-tdr}
\def\svnVersion{6d70f0c}\def\svnDate{2021/09/24}\def\cmsCernNoTag{CERN-EP-2021-091}\def\cmsCernDate{\today}\def\cmsMessage{Published in the Journal of High Energy Physics as \href{http://dx.doi.org/10.1007/JHEP10(2021)176}{\doi{10.1007/JHEP10(2021)176}.}}
\begin{document}\cmsNoteHeader{SMP-20-009}

\newcommand{\zjets}{\ensuremath{\PZ $+$ \text{jets}}\xspace}
\newcommand{\wjets}{\ensuremath{\PW $+$ \text{jets}}\xspace}
\newcommand{\wdijet}{\ensuremath{\PW $+$ \text{dijet}}\xspace}
\newlength\cmsTabSkip\setlength{\cmsTabSkip}{1ex}

\cmsNoteHeader{SMP-20-009} 
\title{Study of \PZ boson plus jets events using variables sensitive to double-parton scattering in \texorpdfstring{$\Pp\Pp$}{pp} collisions at 13\TeV}

\date{\today}

\abstract{Double-parton scattering is investigated using events with a \PZ boson and jets. The \PZ boson is reconstructed using only the dimuon channel. The measurements are performed with proton-proton collision data recorded by the CMS experiment at the LHC at $\sqrt{s}=13\TeV$, corresponding to an integrated luminosity of 35.9\fbinv collected in the year 2016. Differential cross sections of $\PZ$+ $\geq$1 jet and $\PZ$+ $\geq$2 jets are measured with transverse momentum of the jets above 20\GeV and pseudorapidity $\abs{\eta}<2.4$. Several distributions with sensitivity to double-parton scattering effects are measured as functions of the angle and the transverse momentum imbalance between the \PZ boson and the jets. The measured distributions are compared with predictions from several event generators with different hadronization models and different parameter settings for multiparton interactions. The measured distributions show a dependence on the hadronization and multiparton interaction simulation parameters, and are important input for future improvements of the simulations.}

\hypersetup{%
pdfauthor={CMS Collaboration},%
pdftitle={Study of Z boson plus jets events using variables sensitive to double-parton scattering in pp collisions at 13 TeV},%
pdfsubject={CMS},%
pdfkeywords={CMS, Z boson, jets}}

\maketitle 
\section{Introduction}

Proton-proton ($\Pp\Pp$) collisions at high energies result in many events with a \PZ boson produced in association with jets at large transverse momentum \pt. The \zjets process is an important background for many standard model (SM) measurements and searches for physics beyond the SM. Measurements of differential cross sections for \PZ boson production in association with jets can be used to test models of initial-state radiation, and multiparton interactions (MPI). Monte Carlo (MC) models are used to simulate collision processes and play a crucial role in understanding background for the new physics searches. The description of MPI is an important part of these simulations. However, MPI can not be completely described by perturbative quantum chromodynamics (QCD), so this requires a phenomenological description involving parameters that must be tuned with the help of data. 

The ATLAS, CMS, and LHCb experiments have reported various measurements of \zjets processes at centre-of-mass energies of 7\TeV~\cite{Aad:2013ysa,ATLAS:2011wea,Chatrchyan:2011ne,Khachatryan:2014zya,Chatrchyan:2013tna,LHCb:2013mpk}, 8\TeV~\cite{Khachatryan:2015ira,Khachatryan:2016crw,LHCb:2016nhs}, and 13\TeV~\cite{Sirunyan:2018cpw,Aad:2020gfi}. \PZ boson production in association with jets is a process with a clean experimental signature and is well understood theoretically. In this paper, measurements of \zjets events are performed to explore observables sensitive to the presence of MPI. A possible contribution could come from the simultaneous occurrence of two parton-parton interactions and is usually, when the involved scale is large, called double-parton scattering (DPS). This analysis explores new signatures sensitive to DPS in events with a \PZ boson and one or more jets, and compares the data with MC simulations, using a variety of event generators simulating \zjets processes with different MPI and hadronization models. 

The ATLAS, ALICE, CMS, and LHCb Collaborations have previously reported measurements of DPS in various topologies, such as {\PW}{\PW}~\cite{CMS-FSQ-16-005}, \wjets~\cite{sigeff2,sigeff1}, 4 jets~\cite{atlas:dps2018}, \PJGy production~\cite{Aad:2014kba,LHCb:2011kri, LHCb:2016wuo, ALICE:2012pet}, double charm production~\cite{LHCb:2012aiv,LHCb:2020jse}, but not in the \zjets topology.

\begin{figure}[htbp]
\centering
\includegraphics[width=0.42\textwidth]{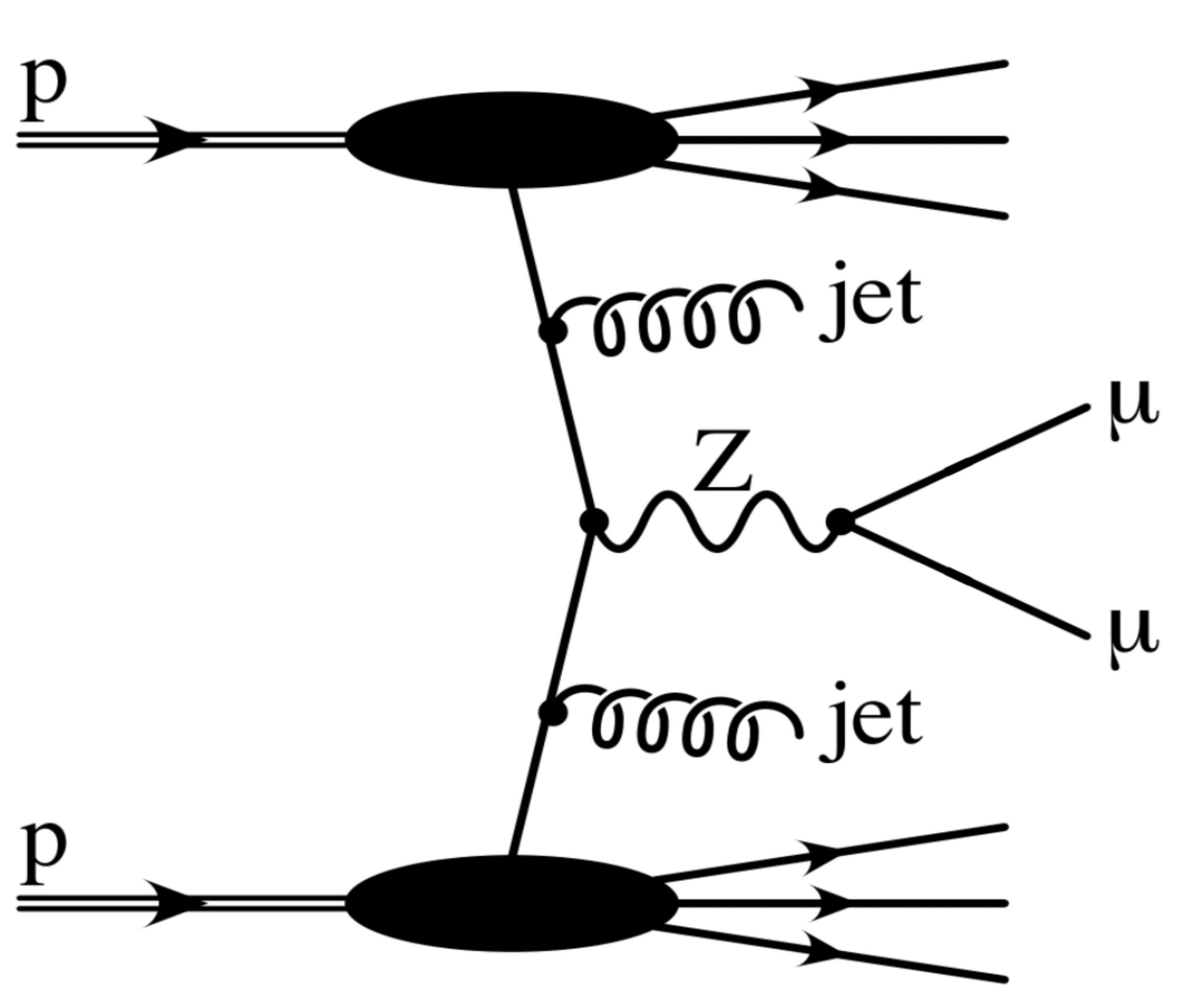}\label{fig:SPS}
\includegraphics[width=0.42\textwidth]{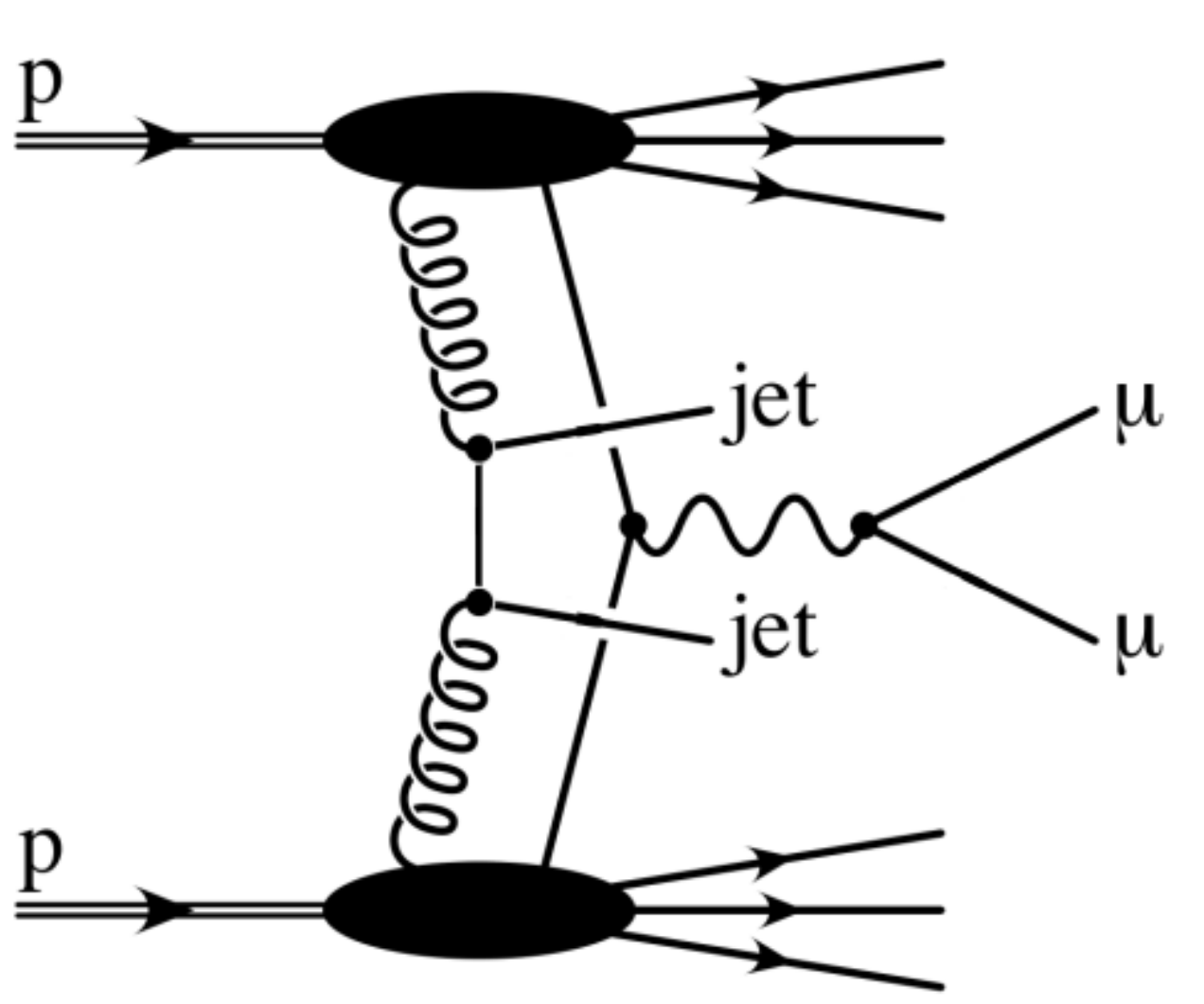}\label{fig:DPS}
 \caption{Typical diagrams for $\PZ$+$2$ jet production in a single (left) and double-parton (right) scattering process.}
\label{fig:SPSDPS}
\end{figure}

Figure~\ref{fig:SPSDPS} shows typical diagrams of single parton scattering (SPS) and DPS production of $\PZ$+$2$ jet events. In SPS, the \PZ boson decaying into two muons and the two jets come from the same parton-parton interaction, whereas in the case of DPS, the \PZ boson and the two jets originate from two independent interactions.

Events are categorized as $\PZ$+ $\ge$1 jet and $\PZ$+ $\ge$2 jets topologies, and the corresponding integrated cross sections are measured in a fiducial region. For consistency with previous DPS measurements at 7\TeV~\cite{sigeff2,sigeff1}, jets are required to have a \pt threshold of 20\GeV. Decreasing the jet \pt threshold would increase the DPS contribution, but will also lead to a significant increase in the experimental uncertainties related to jet identification. With this threshold of 20\GeV, the DPS contribution is measured to be around 5.5\% at 7\TeV~\cite{sigeff1} and, as estimated from simulation, at 13\TeV it is expected to increase by 20--30\% because of the increased parton density. The differential cross sections are measured as functions of various observables based on the azimuthal ($\phi$, in radians) separation and the \pt balance between the \PZ boson and jets, as well as the \pt balance between two jets in case of $\PZ$+ $\ge$2 jets events~\cite{sigeff2, sigeff1, Abazov:2009gc}. Since the two interactions in DPS are largely uncorrelated in the phase space studied, the shape of the distribution of these observables is expected to differ from that seen in SPS. Distributions of these observables normalized to the integrated cross section are measured because they have a lower systematic uncertainty due to the cancellation of the uncertainties correlated among bins. Previously measured distributions were used to extract parameters for DPS modelling in MC event generators~\cite{Sirunyan:2019dfx,Khachatryan:2015pea}.

The measurement is performed using events where the \PZ boson decays into two oppositely charged muons. The $\Pp\Pp$ collision data used in the analysis were collected in 2016 at $\sqrt{s}=13\TeV$, corresponding to an integrated luminosity of 35.9\fbinv. Tabulated results are provided in the HEPDATA~\cite{HEPData}. 

\section{The CMS detector}\label{sec:CMS}

The central feature of the CMS apparatus is a superconducting solenoid of 6\unit{m} internal diameter, providing a magnetic field of 3.8\unit{T}. Within the solenoid volume are a silicon pixel and strip tracker covering a pseudorapidity region of $\abs{\eta}<2.5$, a lead tungstate crystal electromagnetic calorimeter (ECAL), and a brass and scintillator hadron calorimeter, each composed of a barrel and two endcap sections. Forward calorimeters, made of steel and quartz fibers, extend the $\eta$ coverage provided by the barrel and endcap detectors to $\abs{\eta}<5.0$. 
Muons are measured in the range $\abs{\eta}<2.4$, with detection planes made using three technologies: drift tubes, cathode strip chambers, and resistive plate chambers~\cite{Sirunyan:2018}. 

Events of interest are selected using a two-tiered trigger system. The first level, composed of custom hardware processors, uses information from the calorimeters and muon detectors to select events at a rate of about 100\unit{kHz} within a latency of less than 4\mus~\cite{Sirunyan:2020zal}. The second level, known as the high-level trigger, consists of a farm of processors running a version of the full event reconstruction software optimized for fast processing, and reduces the event rate to around 1\unit{kHz} before data storage~\cite{Khachatryan:2016bia}.

  A more detailed description of the CMS detector, together with a definition of the coordinate system used and the relevant kinematic variables, is reported in Ref.~\cite{Chatrchyan:2008zzk}.

\section{Simulated event samples}\label{sec:samples}

MC event generators are used to simulate the signal and background contributions. These MC samples are employed to optimize the event selection, to validate simulated event samples, to estimate some background sources, and to extract the unfolding response matrices used to correct for detector effects in the measured distributions. In addition, MC generators with different MPI and hadronization models are used to compare with the measured distributions and to evaluate systematic uncertainties related to the model dependence.  

The \zjets events are simulated with \MGvATNLO version 2.2.2~\cite{Alwall:2007fs} (denoted MG5\_aMC). The calculation includes matrix elements (MEs) computed at next-to-leading-order (NLO) in perturbative QCD for the process $\Pp\Pp \to \PZ + N\ \text{jets}$, $N = 0$, 1, or 2. The sample is generated using the NNPDF 3.0 NLO parton distribution function (PDF) set~\cite{Ball:2014uwa}. An alternative simulation uses MG5\_aMC to calculate leading-order (LO) MEs for $\Pp\Pp \to \PZ + N\ \text{jets}$, $N = 0, \ldots, 4$, and the NNPDF 3.0 LO PDF set. 

The \zjets process is also simulated with \SHERPA~v2.2~\cite{Bothmann:2019yzt} with up to two additional parton emissions at NLO accuracy or up to four additional parton emissions at LO accuracy. The merging with the \SHERPA parton shower is done via the MEPS@NLO prescription~\cite{Hoeche:2012yf,Gehrmann:2013,Hoeche:2014rya} using the five-flavour-number scheme, with a matching scale of 20\GeV. 
The NNPDF3.0 next-to-NLO (NNLO) set~\cite{Ball:2014uwa} and a dedicated set of tuned parton shower parameters~\cite{Bothmann:2019yzt} developed by the \SHERPA authors are used. 

Top quark-antiquark pair (\ttbar) production, which forms the dominant background, is generated at NLO by MG5\_aMC. The diboson ({\PW}{\PW}, {\PW}{\PZ}, {\PZ}{\PZ}) background processes are simulated at LO using {\PYTHIA}8 v8.212~\cite{Sjostrand:2014zea}, and \POWHEG 2.0 is used for the simulation of the single top quark processes ($s$-channel, $t$-channel, and {\PQt}{\PW}). The prediction for \ttbar production is normalized to NNLO in QCD including resummation of next-to-next-to-leading logarithmic soft-gluon terms with \textsc{Top++} 2.0~\cite{Czakon:2013goa,Baernreuther:2012ws,Beneke:2011mq,Cacciari:2011hy,Czakon:2012zr,Czakon:2012pz,Czakon:2011xx}. Diboson and single top quark background predictions are normalized to NLO~\cite{Campbell:2011bn} and NNLO~\cite{Kidonakis:2013zqa} cross sections, respectively. 

All samples, except those based on \SHERPA, use {\PYTHIA}8 to model the initial- and final-state parton showers and hadronization with the CUETP8M1~\cite{Khachatryan:2015pea} or CUETP8M2T4~\cite{CMS:2016kle} tune. The CUETP8M1 tune includes the NNPDF 2.3~\cite{Ball:2012cx} LO PDF set with the strong coupling $\alpS (m_{\PZ})$ set to 0.1365 for space- and time-like shower simulation. The CUETP8M2T4 tune is based on the CUETP8M1 tune, which includes the NNPDF30\_lo\_as\_0130 PDF set, but uses a lower value of $\alpS = 0.1108$ for the initial-state radiation component of the parton shower. Matching between the ME generators and the parton shower is done using the \kt-MLM scheme~\cite{Alwall:2011uj,Alwall:2008qv} with the matching scale set at 19\GeV for the LO MG5\_aMC sample, and the FxFx~\cite{Frederix:2012ps} scheme with the matching scale set to 30\GeV for the NLO MG5\_aMC events. 

Generated events are processed through a full \GEANTfour--based~\cite{Agostinelli:2002hh} CMS detector simulation and trigger emulation. The simulated samples include the effects of multiple interactions in each bunch crossing, referred to as pileup. The simulated events are reconstructed with the same algorithm used for the data. 

In addition to the \zjets MG5\_aMC samples with the CUETP8M1 tune described above, the measurements are compared with simulations using different {\PYTHIA}8 tunes, such as CP5~\cite{Sirunyan:2019dfx}, CP5 without MPI, and CDPSTP8S1-Wj~\cite{Khachatryan:2015pea}. The CP5 tune uses NNPDF3.1 PDF set at NNLO, with \alpS values of 0.118, and running according to NLO evolution. The CP5 tune is chosen since it is the standard tune obtained by fitting a large number of 1.96, 7, and 13\TeV measurements sensitive to soft and semihard multipartonic interactions~\cite{Sirunyan:2019dfx}. The \wdijet DPS tune, CDPSTP8S1-Wj is derived from the parameters of {\PYTHIA}8 tune 4C, with a variation of the impact parameter dependence, \ie matter overlap function, which is the convolution of the matter distributions of the two incoming hadrons. The DPS simulation in MC is usually quantified in terms of a parameter known as effective cross section $\sigma_\text{eff}$. In {\PYTHIA}8, the value of $\sigma_\text{eff}$ is calculated by dividing the nondiffractive (ND) cross section by the so-called ``enhancement factor", which depends on the parameters of the overlap matter distribution function and the limiting value of \pt at a reference energy~\cite{Sjostrand:1987su}. For central collisions, the enhancement factor tends to be large, translating to a lower value of $\sigma_\text{eff}$ and larger DPS contribution. For peripheral interactions, enhancement factors are small, giving large values of $\sigma_\text{eff}$ and a small DPS contribution. 

The \zjets events calculated at NLO with MG5\_aMC are also interfaced with {\HERWIG}7, applying the CH3 tune for the underlying event description, hadronization, and showering~\cite{Bahr:2008pv,Bellm:2015jjp,Bellm:2017bvx,Sirunyan:2020pqv}. The tune is derived by fitting measurements from $\Pp\Pp$ collision data collected by the CMS experiment at $\sqrt{s} = 0.9$, 7, and 13\TeV. The CH3 tune makes use of the NNPDF3.1 LO PDF set with $\alpS = 0.130$ for the simulation of MPI, but the NNLO PDF set with $\alpS = 0.118$ for all other components. The cross sections of \zjets events simulated with different MC event generators and configurations are normalized to NNLO calculations with \FEWZ v3.1~\cite{Melnikov:2006kv}.
Table~\ref{tab:MC} summarizes all the event generators, PDF sets, and tunes used to produce both signals (including alternative samples), and the background processes.

\begin{table}[]
\centering
\topcaption{Summary of event generators with their k-factors (ratio of NNLO to NLO/LO cross section), PDF sets, and tunes used to produce both the signal and background event samples.}
\resizebox{\textwidth}{!}{%
\begin{tabular}{lccc}
\hline
Process name     & Event generator (k-factor)                  & Tune         & PDF set                                                                                     \\ \hline
\zjets           & MG5\_aMC (NLO) + {\PYTHIA}8 (1.07)         & CUETP8M1     & \begin{tabular}[c]{@{}l@{}}NNPDF 3.0 NLO for matrix element\\ NNPDF 2.3 LO for tune\end{tabular} \\
\zjets           & MG5\_aMC (LO) + {\PYTHIA}8 (1.24)          & CUETP8M1     & \begin{tabular}[c]{@{}l@{}}NNPDF 3.0 LO for matrix element\\ NNPDF 2.3 LO for tune\end{tabular} \\
\zjets           & SHERPA (NLO) (0.97)                     & \NA            & NNPDF 3.0 NNLO                                                                                   \\
\zjets           & MG5\_aMC (NLO) + {\PYTHIA}8 (1.07)         & CP5          & \begin{tabular}[c]{@{}l@{}}NNPDF 3.0 NLO for matrix element\\ NNPDF 3.1 NNLO for tune\end{tabular} \\
\zjets           & MG5\_aMC (NLO) + {\PYTHIA}8 MPI-OFF (1.07) & CP5          & \begin{tabular}[c]{@{}l@{}}NNPDF 3.0 NLO for matrix element\\ NNPDF 3.1 NNLO for tune\end{tabular} \\
\zjets           & MG5\_aMC (NLO) + {\PYTHIA}8 (1.09)         & CDPSTP8S1-Wj & \begin{tabular}[c]{@{}l@{}}NNPDF 3.0 NLO for matrix element\\ CTEQ6L1 LO for tune\end{tabular}   \\
\zjets           & MG5\_aMC (NLO) + {\HERWIG}7 (1.02)         & CH3          & \begin{tabular}[c]{@{}l@{}}NNPDF 3.0 NLO for matrix element\\ NNPDF 3.1 LO/NNLO for tune\end{tabular} \\ 
\ttbar & MG5\_aMC (NLO) + {\PYTHIA}8 (\NA)         & CUETP8M2T4   & \begin{tabular}[c]{@{}l@{}}NNPDF 3.0 NLO for matrix element\\ NNPDF30\_lo\_as\_0130 for tune\end{tabular} \\
Diboson          & {\PYTHIA}8 (LO) (\NA)                     & \NA            & NNPDF 2.3 LO                                                                                   \\
Single top       & \POWHEG (NLO) + {\PYTHIA}8 (\NA)           & CUETP8M1     & \begin{tabular}[c]{@{}l@{}}NNPDF 3.0 NLO for matrix element\\ NNPDF 2.3 LO for tune\end{tabular}
\end{tabular}\label{tab:MC}
}
\end{table}

\section{Event selection}\label{sec:objects}

The particle flow (PF) algorithm~\cite{pf} is used to reconstruct and identify individual particle candidates in an event, with an optimized combination of information from the various elements of the CMS detector. Energy deposits are measured in the calorimeters and charged particles are identified in the central tracking and muon systems.

Events are selected with a single-muon trigger requiring at least one isolated muon candidate with $\pt>24\GeV$. The primary vertex is the vertex candidate with the largest value of summed physics-object $\pt^2$. The physics objects are the track-only jets, clustered using the jet finding anti-\kt algorithm~\cite{Cacciari:2008gp,Cacciari:2011ma} with the tracks assigned to candidate vertices as inputs, and the associated missing \pt, which is the negative vector \pt sum of all physics objects. Events are required to have at least two oppositely charged muons  with $\pt>27\GeV$ and $\abs{\eta}<2.4$. These muons are reconstructed by combining information from the inner tracker and the muon detector subsystems. The muon candidates are required to satisfy identification criteria based on the number of hits in each detector, the quality of the muon-track fit, and the consistency with the primary vertex, which is imposed by requiring the longitudinal and transverse impact parameters are less than 0.5 and 0.2\cm, respectively. 
The efficiency to reconstruct and identify the muons is greater than 96\%. Matching muon candidates to tracks measured in the silicon tracker results in a relative \pt resolution for muons with $20<\pt<100\GeV$ of 1\% in the barrel and  3\% in the endcaps.  

To suppress the multijet background,  muons are required to be isolated. The relative isolation variable, $\mathrm{I}_\text{rel}$, for muons is defined as:
\begin{linenomath}
\begin{equation}
\mathrm{I}_\text{rel} = \frac{[\sum \pt^{{\, \text{charged}}} + \max(0., \sum \ET^{{\, \text{neutral}}} + \sum \ET^{\gamma} - 0.5\sum \pt^\mathrm{ PU})]}{\pt^{\PGm}}.
\end{equation} 
\end{linenomath}
Here $\sum \ET^{\smash[b]{\, \text{neutral}}}$ and $\sum \ET^{\gamma}$ are the transverse energy sums of neutral hadrons and photons, respectively, within a cone of radius $\Delta R = \sqrt{\smash[b]{(\Delta\eta)^2+(\Delta\phi)^2}} = 0.4$ around the muon track. The quantity $\sum \pt^{\smash[b]{\, \text{charged}}}$ represents the \pt sum of the charged hadrons in the same cone around the muon associated with the selected vertex. Finally, $\sum\pt^\mathrm{PU}$ is the \pt sum of the charged hadrons in the same cone around the muon not associated with the selected vertex. A muon is considered isolated if $\mathrm{I}_\text{rel}<0.2$. There are small residual differences in the trigger, identification and isolation efficiencies between data and simulation, which
are measured using the ``tag-and-probe" method~\cite{wz}, and included by applying scale factors to simulated events~\cite{Sirunyan:2018}.

To select \PZ boson candidate events, the invariant mass of the two oppositely charged muons with highest \pt is required to be close to the \PZ boson mass, $70<m_{\PGmp\PGmm}<110\GeV$.  
The \PZ boson candidate is required to be accompanied by at least one jet with $\pt>20\GeV$ and $\abs{\eta}<2.4$. The overlap between the muons from the \PZ boson decay and the jets is removed by requiring a minimum $\Delta R$ distance of 0.4 between them.

Jets are clustered from PF candidates using the infrared- and collinear-safe anti-\kt algorithm~\cite{Cacciari:2008gp} with a distance parameter of 0.4, as implemented in the \FASTJET package \cite{Cacciari:2011ma}.
The jet momentum is determined as the vectorial sum of all particle momenta in the jet, and is found from simulation to be, on average, within 5 to 10\% of the true momentum over the entire \pt spectrum and detector acceptance. Pileup can contribute additional tracks and calorimetric energy depositions, increasing the apparent jet momentum. To mitigate the effects of pileup, tracks identified as originating from pileup vertices are discarded, and a factor~\cite{pileups} is applied to correct for the remaining contributions. Jet energy corrections are derived from simulation studies so that the average measured response of jets becomes identical to that of particle-level jets. In situ measurements of the momentum balance in dijet, multijet, photon+jet, and \zjets events with the \PZ boson decaying leptonically are used to correct residual differences between the jet energy scales in data and simulation~\cite{Khachatryan:2016kdb,CMS-DP-2018-028}. The jet energy in simulation is spread to match the resolution observed in data. The jet energy resolution (JER) in data amounts typically to 15--20\% at jet energy of 30\GeV, 10\% at 100\GeV, and 5\% at 1\TeV. Additional selection criteria are applied to remove jets potentially dominated by anomalous contributions from various subdetector components or reconstruction failures~\cite{cms:jp}. Jets identified as being likely to originate from pileup~\cite{CMS-PAS-JME-13-005,CMS-PAS-JME-18-001} are also removed by using a pileup jet ID discriminator for jets with $20<\pt\leq50\GeV$. No pileup jet ID is applied to jets with $\pt\geq50\GeV$, since here the pileup event contribution is negligible~\cite{CMS-PAS-JME-18-001}.

The selected events correspond to a sample of \zjets events with a background of 2--5\%, which is subtracted from the data before the unfolding.
The dominant contribution is the production of \ttbar pairs. 
The simulation of \ttbar production is validated with data in a $\ttbar \to \Pe\PGm + \mathrm{X}$ control region. The \ttbar control region is constructed using data events by requiring an oppositely charged electron-muon pair with at least one jet. The energy of electrons is determined from a combination of the electron momentum at the primary interaction vertex as determined by the tracker, the energy of the corresponding ECAL cluster, and the energy sum of all bremsstrahlung photons spatially compatible with originating from the electron track~\cite{Sirunyan:2020ycc}. The longitudinal and transverse impact parameters for barrel (endcap) are required to be less than 0.10 (0.20) and 0.05 (0.10)\cm, respectively. Electrons are required to have $\pt>27\GeV$ and $\abs{\eta}<2.4$. The selection criteria for muons and jets are the same as discussed above. The invariant mass of the oppositely charged $\Pe\PGm$ pair is required to lie within the range of $70<m_{\Pe\PGm}<110\GeV$. The difference in the data to simulation comparison in this control region is included as a part of the systematic uncertainty.

\section{Observables and the unfolding} \label{sec_Unf}

Events are categorized in the $\PZ$+ $\geq$1 jet and $\PZ$+ $\geq$2 jets subsets, and the differential cross sections are measured as function of the following observables:

\begin{enumerate}

\item {$\PZ$+ $\geq$1 jet events}

\begin{itemize}
\item Azimuthal angle  between the \PZ boson and the highest \pt (leading) jet: $\Delta \phi (\PZ, j_1)$.
For SPS events, the \PZ boson and the leading jet will balance each other, hence this observable will peak around $\pi$, whereas for the DPS process the distribution is expected to be flat because of the absence of a correlation between the \PZ boson and the leading jet produced from the two different scatterings.

\item Relative \pt imbalance between the \PZ boson and leading jet is
\begin{equation}
 \Delta_\text{rel}\pt(\PZ,j_1) = \frac{\abs{{\vec\pt(\PZ)} + {\vec\pt(j_1)}}}{\abs{{\vec\pt(\PZ)}} + \abs{{\vec\pt(j_1)}}}.
\end{equation}
This observable is expected to be close to zero for SPS events, whereas in the case of DPS this observable will have higher values.
\end{itemize}

\item {$\PZ$+ $\geq$2 jets events}

\begin{itemize}
\item Azimuthal angle between the \PZ boson and dijet system: $\Delta \phi$(\PZ, dijet).
Here, dijet means the resulting three-momentum of the leading and subleading jets.
For SPS events, the dijet system \pt will balance the \PZ boson \pt, therefore, this observable will peak around $\pi$. In the case of DPS production, the distribution is expected to be flat since the \PZ boson and the two jets are originating from two independent scatterings.

\item Relative \pt imbalance between the \PZ boson and dijet system is
\begin{equation}
 \Delta_\text{rel}\pt(\PZ,\text{dijet}) = \frac{\abs{{\vec\pt(\PZ)} + {\vec\pt(\text{dijet})}}}{\abs{{\vec\pt(\PZ)}} + \abs{{\vec\pt(\text{dijet})}}}.
 \end{equation}
For SPS events this observable is expected to be 0, whereas for DPS events this observable will have higher values. 
 
\item Relative \pt imbalance between the leading ($j_1$) and subleading ($j_2$) jets is
\begin{equation}
\Delta_\text{rel}\pt(j_1,j_2) = \frac{\abs{{\vec\pt(j_1)} + {\vec\pt(j_2)}}}{\abs{{\vec\pt(j_1)}} + \abs{{\vec\pt(j_2)}}}.
\end{equation}
For DPS events, the two jets are expected to balance each other, therefore this observable will be around 0. For SPS events, the two jets are correlated with the \PZ boson and not expected to balance each other.

\end{itemize}

\end{enumerate}

The reconstructed distributions are corrected for the event selection efficiency and detector resolution using an unfolding technique that employs a response matrix to map the reconstructed observables onto the particle-level values. The unfolding is done with 20 detector-level and 10 particle-level $\Delta\phi$ or $\Delta_\text{rel}\pt$ bins. The unfolding is performed using the \textsc{TUnfold} package~\cite{Schmitt:2012kp}, which is based on a least squares fit with a possible Tikhonov regularization term~\cite{Tikhonov:1963}. Since the effect of regularization is minimal on the reported observables, the unfolding is performed without regularization.

The \zjets events simulated with MG5\_aMC + {\PYTHIA}8 with tune CUETP8M1 are used to construct the response matrix. At particle level, events are required to have at least two oppositely charged muons with $\pt>27\GeV$ and $\abs{\eta}<2.4$. The particle-level definition of a muon corresponds to a generator-level muon coming from the \PZ boson decay, ``dressed" by adding the momenta of all photons within $\Delta R<0.1$ around both muon directions to account for the FSR effects~\cite{Collaboration:2267573}. The particle-level jets with $\pt>20\GeV$ and $\abs{\eta}<2.4$ are formed from stable particles ($c\tau > 1\cm$), except neutrinos, using the same anti-\kt jet algorithm as for reconstructed jets. A possible overlap between particle-level jets and a pair of muons from the \PZ boson decay is removed by requiring a minimum distance of 0.4 between them. The distributions are unfolded to the particle level in the fiducial region defined in Table~\ref{tab:fiducial}. 

\begin{table}[h]
\centering
\topcaption{Fiducial selections at particle level.}
\begin{tabular}{lc}
\hline
Object & Selections \\ \hline
Muons (dressed) & $\pt>27\GeV$, $\abs{\eta}<2.4$        \\  
\PZ boson & $70<m_{\PGmp\PGmm}<110\GeV$                   \\ 
At least 1 jet & $\pt>20\GeV$, $\abs{\eta}<2.4$                    \\ 
\end{tabular}%
\label{tab:fiducial}
\end{table}

In simulation, the reconstructed jets and a pair of muons are spatially matched to the corresponding particle-level objects by requiring that they are within $\Delta R$ of 0.4 from one another. Events that have reconstructed objects without matched particle-level objects are included in the background category and are excluded from the sample. 
This contribution includes the events where the selected jets originate from pileup. The contribution of background with no jet at particle level, but at least one jet at the reconstructed level is about 4.1\%. The contribution of background with no jet (1 jet) at particle level and at least 2 jets at the reconstruction level is around 1.1 (4.8)\%.
The simulation of pileup jets is validated in a control region enriched in pileup jets, obtained by inverting the criteria used to reject the pileup jets. 
The simulation describes the data well, within the uncertainties, in the pileup-enriched control region, validating the simulation of background events from pileup. 

Events that have particle-level objects in the fiducial volume, but no matching reconstructed objects, are accounted for with acceptance and efficiency corrections. In addition, there may be events in which the particle-level jet passing the fiducial selection does not lead to a reconstructed level jet that passes the fiducial selection. Events of this type are considered as background at the reconstruction level and are not considered. However, at the generator level, these are genuine signal events missed because of the detector and reconstruction inefficiencies. These missed events are accounted for via a signal acceptance correction. 

Finally, the unfolded distributions are scaled with the inverse of the integrated luminosity to obtain the differential cross section.

\section{Systematic uncertainties} \label{sec_syst}

The measurements have various sources of systematic uncertainties. 

\begin{itemize}

\item{Jet energy resolution and scale}: The effect of the uncertainty in jet energy scale (JES) or JER~\cite{Khachatryan:2016kdb,CMS-DP-2018-028} is evaluated by varying the JES or JER within the associated uncertainty and performing the unfolding procedure with the modified distribution. The variations of JES corrections within their uncertainties change the differential cross sections by 2--8\%, whereas the area-normalized distributions are affected by up to 4\%. The variations of JER corrections within their uncertainty change the differential cross section distributions by 1--7\% and area-normalized distributions up to 5\%.

\item{Pileup jet identification:} Simulated events are corrected for the differences in the jet identification efficiency between data and simulated events. The uncertainties in these corrections affect the measurements by up to 1--2\% for differential cross section distributions and less than 0.5\% for area-normalized distributions.

\item {Closure uncertainty}: The effect of model dependence is evaluated by comparing unfolded results obtained using response matrices constructed with the MG5\_aMC + {\PYTHIA}8 with tune CUETP8M1 and \SHERPA generators having different ME and parton showering, as discussed in Section~\ref{sec:samples}. The calculated uncertainty is then symmetrized. 

The effect of scale uncertainties is estimated using a set of generator weights that correspond to variations of renormalization ($\mu_{F}$) and factorization ($\mu_{R}$) scales up and down by factors of 2 from their nominal values, excluding the pair of extreme variations. The unfolded distributions are obtained for all such combinations and their envelope is quoted as the uncertainty. 

The uncertainty in the PDFs is estimated using the 100 replicas of the NNPDF 3.0 PDF set~\cite{Butterworth:2015oua}. The unfolded distributions are reproduced using the weights of the replicas and a standard deviation is computed on a bin-by-bin basis~\cite{Butterworth:2015oua}. 

These sources, added in quadrature, affect the differential cross section by 1--5\% and area-normalized distributions up to 1--4\% in the case of $\PZ$+ $\geq$1 jet events, whereas the effect is within 9\% for the differential cross section and up to 7\% for the area-normalized distribution in the case of $\PZ$+ $\geq$2 jet events.

\item{Integrated luminosity}: It is determined with a 2.5\%~\cite{CMS-LUM-17-003} uncertainty for differential cross section distributions, but completely cancels in the area normalized distributions.

\item{Pileup weighting}: The distribution of the mean number of interactions per bunch crossing of the simulated samples is weighted to match that of the data. The uncertainty related to pileup weighting is estimated by varying the total inelastic cross section by $\pm$4.6\%~\cite{Sirunyan:2018nqx}. The effect is negligible for both the differential and area-normalized distributions. After the pileup weighting, the vertex multiplicity in simulation shows overall good agreement with data, but there is an overestimation of around 40\% for high vertex multiplicities. 
To investigate the effect of this residual discrepancy, the simulated events are additionally corrected to reproduce the vertex multiplicity distributions observed in data.
The data are unfolded with the weighted response matrix, and the results are compared with the unfolded results without weighting.
These sources, added in quadrature, affect the differential cross section by 1.0--1.5\% and area-normalized distributions up to 1\%.

\item{Muon selection}: The systematic uncertainty related to various muon selection criteria such as muon identification, isolation and trigger scale factors is less than 1\% for both differential cross section and area-normalized distributions.

The effect of the muon momentum corrections on the measurement is very small ($<$0.1\%), therefore no additional systematic uncertainty is assigned.
 
\item{Background modelling}: There is a small contribution from \ttbar events, whereas the contribution from other processes such as dibosons, \wjets, and QCD multijet production is smaller than 1\%. To calculate the systematic uncertainty related to the simulated background contribution, the cross sections of the background samples are varied by their uncertainties. The systematic uncertainty related to the \ttbar process is obtained from the differences between data and simulation in a $\ttbar \to \Pe\PGm + \mathrm{X}$ control region. The variation of the background contribution within the uncertainties affects the differential cross section less than than 0.2\% for $\PZ$+ $\geq$1 jet and less than 0.6\% for $\PZ$+ $\geq$2 jets events. The effect of this variation on the area-normalized distributions is less than 0.2\%.

\end{itemize}

Tables~\ref{tab:expunc} and~\ref{tab:expunc_SC} summarize the effect of various systematic uncertainties for the differential cross section and the normalized distributions. These systematic uncertainties are considered uncorrelated and are added in quadrature. 

\begin{table}[htbp]
\centering
\topcaption{Uncertainty sources and their effect on the differential cross section distributions.}
\resizebox{\columnwidth}{!}{
\begin{tabular}{lccccc}
\hline
Observable/Uncertainty & $\Delta \phi (\PZ, j_1)$  &   $\Delta_\text{rel}\pt(\PZ,j_1)$ & $\Delta \phi$(\PZ, dijet) & $\Delta_\text{rel}\pt(\PZ,\text{dijet})$ & $\Delta_\text{rel}\pt(j_1,j_2)$ \\ \hline
JES  & 2.7--7.5\% & 2.4--7.4\%  & 4.9--7.9\% & 4.5--8.4\% &4.4--7.3\% \\
JER  & 0.9--6.6\% & 1.4--5.8\%  & 1.2--7.2\% & 2.1--5.1\% & 1.1--4.2\% \\
Pileup jet identification & 1.3--1.7\% & 0.9--1.6\%  & 1.7--2.1\% & 1.6--2.1\% & 1.7--2.3\% \\
Integrated luminosity & 2.5\% & 2.5\% & 2.5\% & 2.5\% & 2.5\% \\
Pileup modelling & 0.1--0.7\% & 0.2--1.0\%  & 0.2--1.4\% & 0.4--1.4\% & 0.8--1.4\% \\
Closure uncertainty & 0.6--4.0\% & 0.8--5.1\%  & 2.7--6.1\% & 2.2--8.7\% & 2.2--8.7\% \\
Muon selection & $<$1.0\% & $<$1.0\%  & $<$1.0\% & $<$1.0\% & $<$1.0\% \\
Background modelling & $<$0.2\% & $<$0.2\%  & $<$0.6\% & $<$0.6\% & $<$0.4\% \\[\cmsTabSkip] 
Total & 4--11\% & 4--10\% & 8--14\% & 8--14\% & 7--11\% \\
\end{tabular}}
\label{tab:expunc}
\end{table}

\begin{table}[htbp]
\centering
\topcaption{Uncertainty sources and their effect on the area-normalized distributions.}
\resizebox{\columnwidth}{!}{
\begin{tabular}{lccccc}
\hline
Observable/Uncertainty & $\Delta \phi (\PZ, j_1)$     & $\Delta_\text{rel}\pt(\PZ,j_1)$ & $\Delta \phi$(\PZ, dijet) & $\Delta_\text{rel}\pt(\PZ,\text{dijet})$ & $\Delta_\text{rel}\pt(j_1,j_2)$ \\ \hline
JES  & 0.1--3.8\% & 0.7--3.7\%  & 0.6--4.0\% & 0.3--2.6\% & 0.3--1.5\% \\
JER  & 0.3--4.6\% & 0.4--4.4\%  & 1.3--4.4\% & 0.2--4.8\% & 0.2--1.7\% \\
Pileup jet identification & 0.1--0.2\% & 0.1--0.2\% & 0.1--0.2\% &  0.1--0.2\% &  0.1--0.4\% \\
Pileup modelling & 0.1--0.5\% & 0.1--0.5\%  & 0.1--1\% & 0.1--0.8\% & 0.2--0.4\% \\
Closure uncertainty & 0.8--2.5\% & 0.9--3.6\%  & 0.3--5.0\% & 0.4--6.7\% & 0.5--3.7\% \\
Muon selection & $<$1.0\% & $<$1.0\%  & $<$1.0\% & $<$1.0\% & $<$1.0\% \\
Background modelling & $<$0.1\% & $<$0.1\%  & $<$0.2\% & $<$0.2\% & $<$0.2\% \\[\cmsTabSkip]
Total & 1--6\% & 1--6\% & 2--7\% & 1--7\% & 1--4\% \\
\end{tabular}}
\label{tab:expunc_SC}
\end{table}

\section{Results} \label{sec_result}

The production cross sections in the fiducial region defined in Table~\ref{tab:fiducial} are measured to be $158.5\pm 0.3\stat\pm 7.0\syst\pm 1.2\thy\pm 4.0\lum$\unit{pb} for $\PZ$+ $\geq$1 jet events and $44.8\pm 0.4\stat\pm 3.7\syst\pm 0.5\thy\pm 1.1\lum$\unit{pb} for $\PZ$+ $\geq$2 jet events.
The measured cross sections are described, within the uncertainties, by different simulations, except for the  MG5\_aMC + {\PYTHIA}8 with CP5 tune MPIOFF and the DPS-specific CDPSTP8S1-WJ tune. The cross section of the DPS-specific tune is predicted to be 10\% higher than the measured cross section. The cross section of CP5 tune without MPI underestimates the measured cross section up to 10\% for $\PZ$+ $\geq$1 jet events and up to 16\% for $\PZ$+ $\geq$2 jet events.
Table~\ref{tab:integrated_cs} summarizes the measured and predicted cross sections for the $\PZ$+ $\geq$1 jet and $\PZ$+ $\geq$2 jet processes.

\begin{table}[htbp]
\centering
\topcaption{Measured and predicted cross section for $\PZ$+ $\geq$1 jet and $\PZ$+ $\geq$2 jet production. Simulations are normalized to NNLO calculations (from \FEWZ) and reported cross section values are extracted by applying fiducial selections. Predicted cross sections include statistical and theoretical uncertainties added in quadrature for MG5\_aMC + {\PYTHIA}8 with CP5 tune, whereas the other predicted cross sections are reported with only statistical uncertainties.}
\begin{tabular}{llll}
\hline
\multicolumn{2}{l}{Cross section (pb)}                              & $\PZ$+ $\geq$1 Jets & $\PZ$+ $\geq$2 Jets \\ \hline
Measured in data             &                                             & \vtop{\hbox{\strut $158.5\pm 0.3\stat$} \hbox{\strut $\pm 7.0\syst$} \hbox{\strut $\pm 1.2\thy$} \hbox{\strut  $\pm 4.0\lum$\unit{pb}}}  & \vtop{\hbox{\strut $44.8\pm 0.4\stat$} \hbox{\strut $\pm 3.7\syst$} \hbox{\strut $\pm 0.5\thy$} \hbox{\strut $\pm 1.1\lum$\unit{pb}}} \\
Predicted by MC & & & \\
\multirow{4}{*}{MG5\_aMC (NLO)} & \multicolumn{1}{|l}{{\PYTHIA}8, CP5 tune}           & 167.4 $\pm$ 9.7 & 47.0 $\pm$ 3.9  \\  
                        & \multicolumn{1}{|l}{{\PYTHIA}8, CP5 tune MPIOFF} & 143.8 $\pm$ 0.3 & 37.7 $\pm$ 0.2  \\
                        & \multicolumn{1}{|l}{{\PYTHIA}8, CDPSTP8S1-WJ tune} & 178.4 $\pm$ 0.3 & 50.5 $\pm$ 0.2  \\  
                        & \multicolumn{1}{|l}{{\HERWIG}7, CH3 tune}           & 158.3 $\pm$ 1.1 & 44.4 $\pm$ 0.6  \\ 
\multicolumn{2}{l}{MG5\_aMC (LO) + {\PYTHIA}8, CP5 tune}                                    & 161.2 $\pm$ 0.1 & 45.3 $\pm$ 0.1  \\ 
\multicolumn{2}{l}{\SHERPA (NLO+LO)}                                          & 149.8 $\pm$ 0.2 & 41.6 $\pm$ 0.1  \\
\end{tabular}%
\label{tab:integrated_cs}
\end{table}

\begin{figure}[htbp]
\centering
\includegraphics[width=0.48\textwidth]{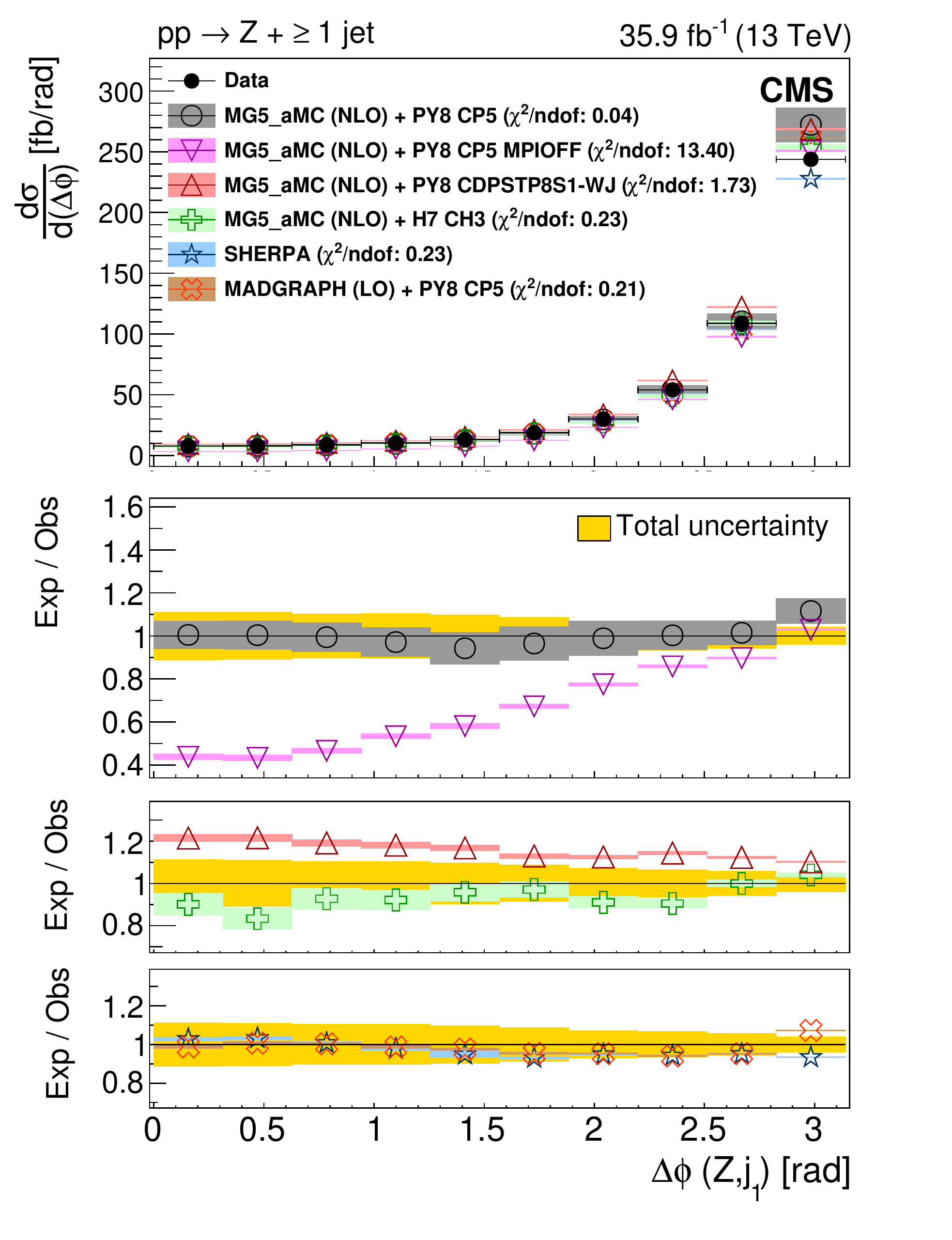}\label{fig:1dataMC_dphi_Z1J_CN_13TeV.pdf}
\includegraphics[width=0.48\textwidth]{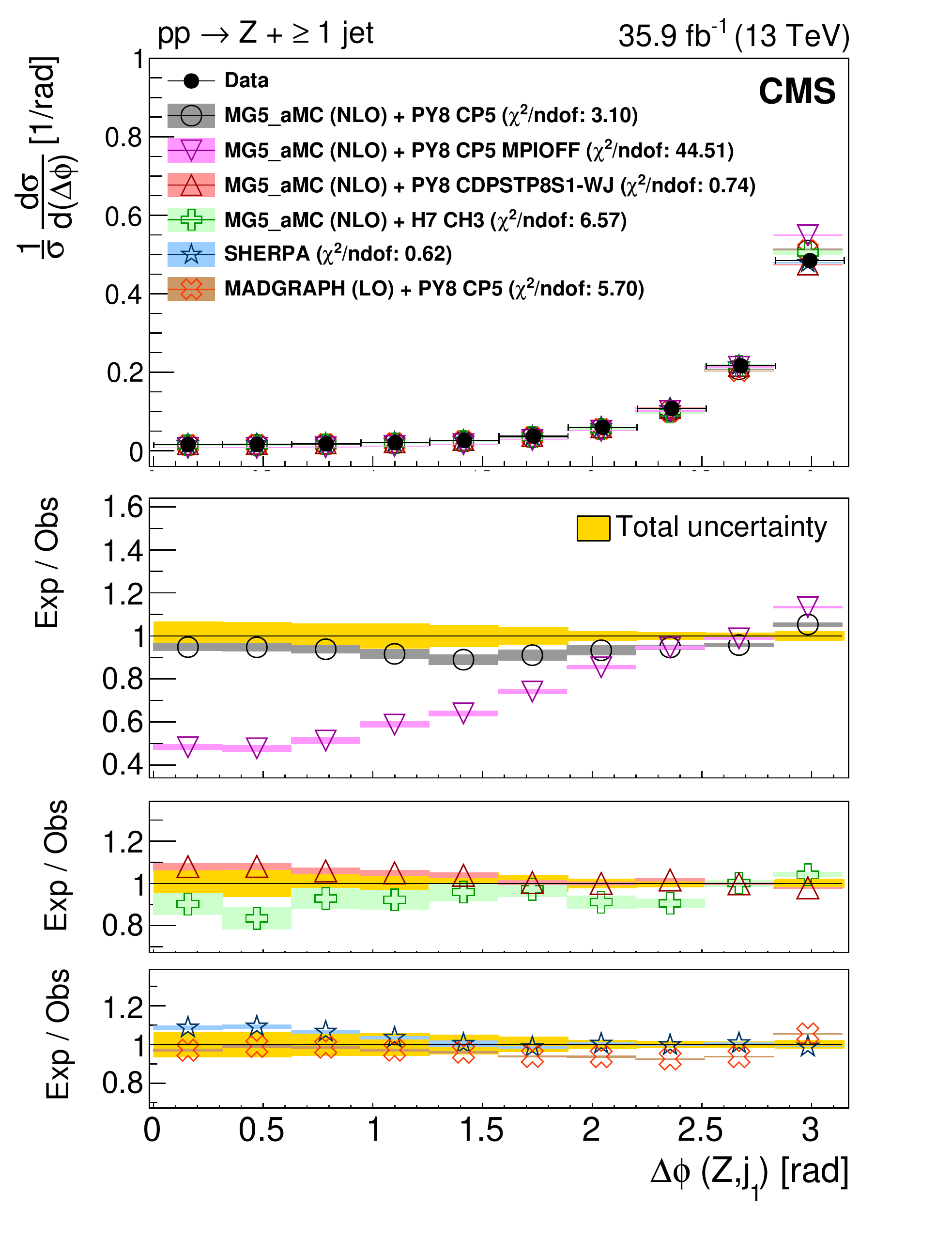}\label{fig:0dataMC_dphi_Z1J_SC_13TeV.pdf}
\caption{Differential cross sections (left) and area-normalized distributions (right) as functions of $\Delta\phi$ between the \PZ boson and the leading jet for $\PZ$+ $\geq$1 jet events. The uncertainties in the predictions are shown as coloured bands around the theoretical predictions including statistical, PDF, scale, and tune uncertainties for the NLO MG5\_aMC + {\PYTHIA}8 (with CP5 tune) and the statistical uncertainty only for the LO MG5\_aMC + {\PYTHIA}8 (with CP5 tune), NLO MG5\_aMC + {\PYTHIA}8 (with CDPSTP8S1-WJ tune, CP5 tune with MPI-OFF), NLO MG5\_aMC + {\HERWIG}7 (with tune CH3), and \SHERPA predictions. In the top panel, the vertical bars on the data points represent statistical uncertainties, whereas in the bottom panels, the total uncertainty in data is indicated by the solid yellow band centred at 1. In the legend, the $\chi^{2}$ per degree of freedom is given to quantify the goodness of fit of the model to the data.}
\label{fig:Z1J_dphi}
\end{figure}

\begin{figure}[htbp]
\centering
\includegraphics[width=0.48\textwidth]{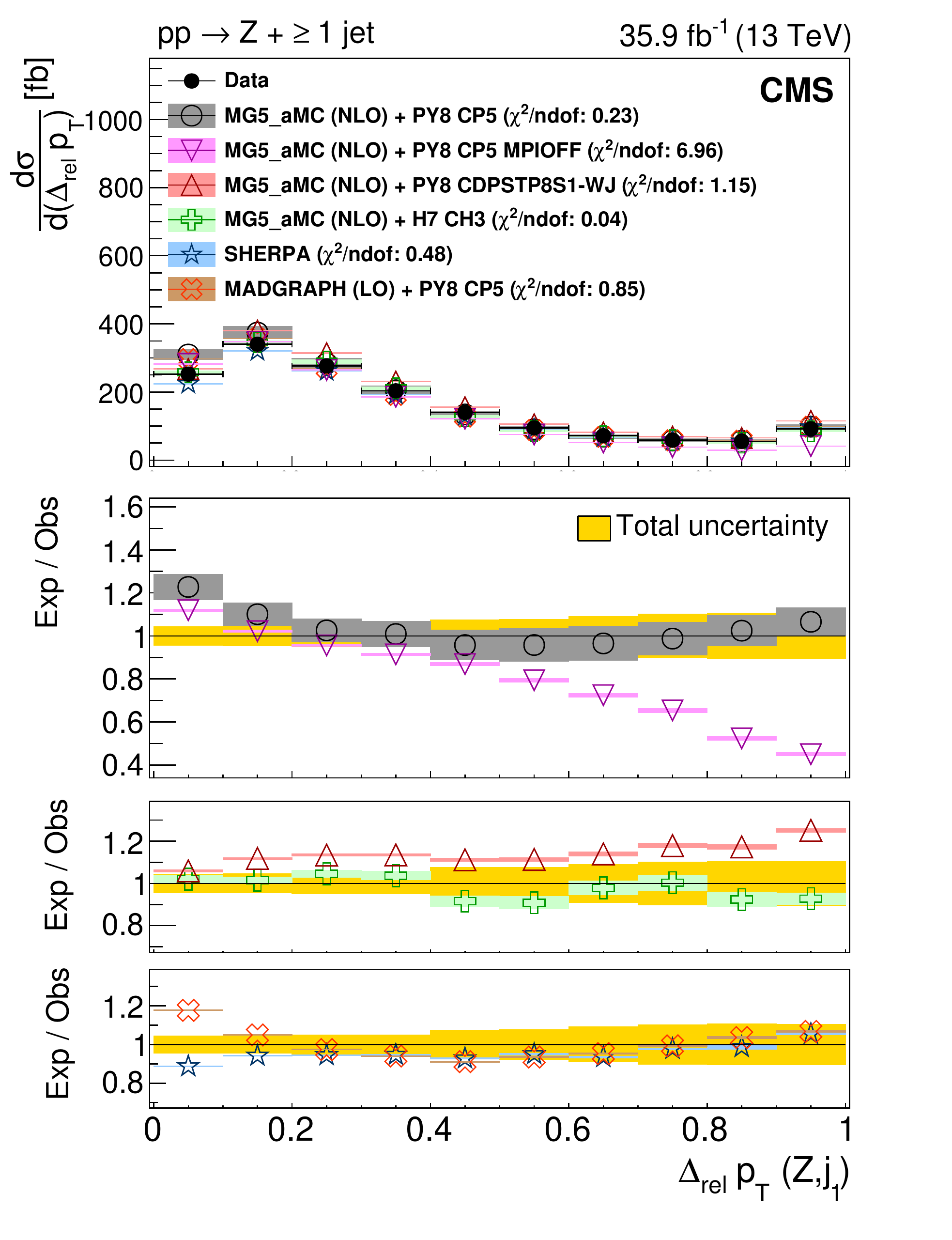}\label{fig:1dataMC_reldpt_Z1J_CN_13TeV.pdf}
\includegraphics[width=0.48\textwidth]{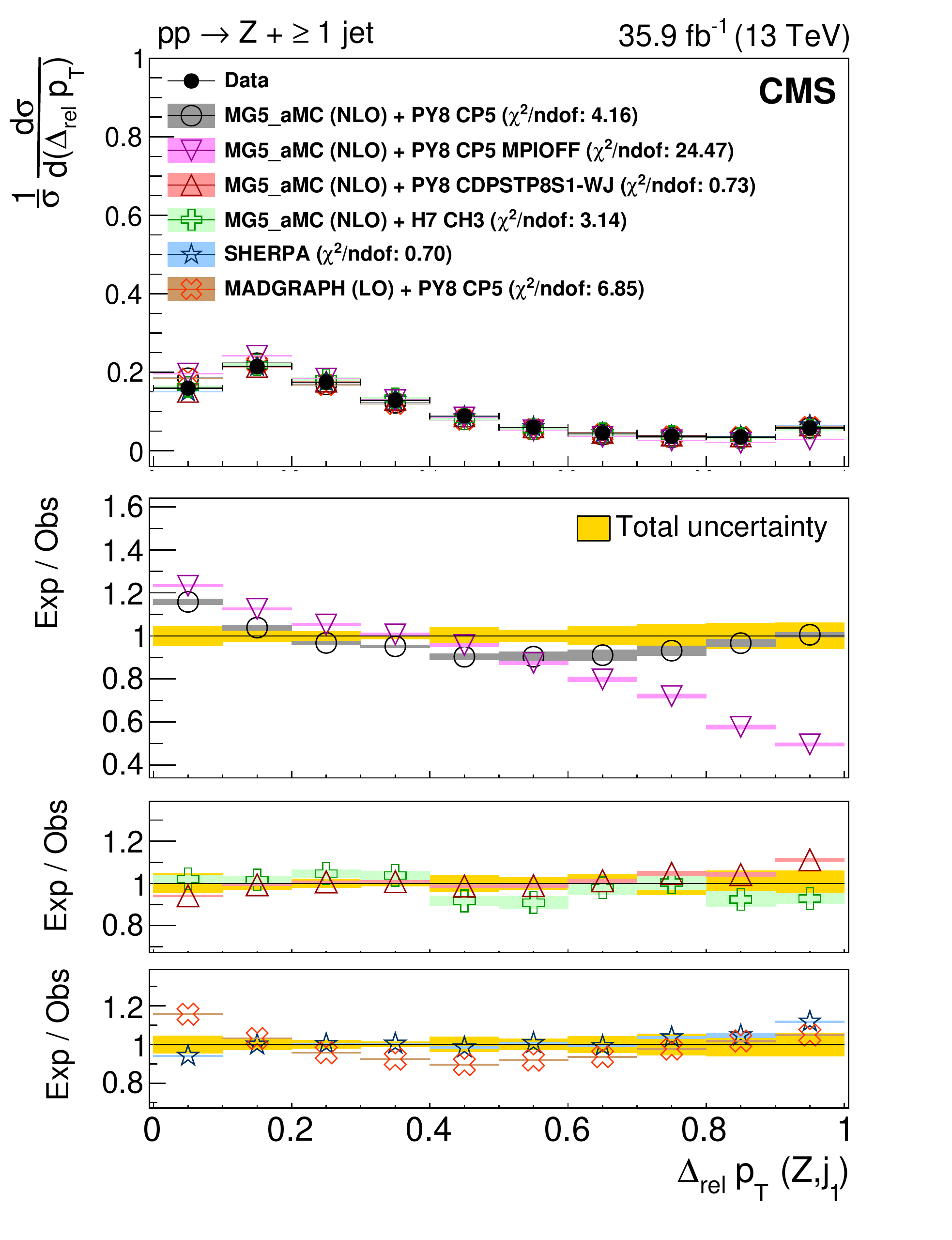}\label{fig:0dataMC_reldpt_Z1J_SC_13TeV.pdf}
\caption{Differential cross sections (left) and area-normalized distributions (right) as functions of the \pt imbalance between the \PZ boson and the leading jet for $\PZ$+ $\geq$1 jet events. The uncertainties in the predictions are shown as coloured bands around the theoretical predictions including statistical, PDF, scale, and tune uncertainties for the NLO MG5\_aMC + {\PYTHIA}8 (with CP5 tune) and the statistical uncertainty only for the LO MG5\_aMC + {\PYTHIA}8 (with CP5 tune), NLO MG5\_aMC + {\PYTHIA}8 (with CDPSTP8S1-WJ tune, CP5 tune with MPI-OFF), NLO MG5\_aMC + {\HERWIG}7 (with tune CH3), and \SHERPA predictions. In the top panel, the vertical bars on the data points represent statistical uncertainties, whereas in the bottom panels, the total uncertainty in data is indicated by the solid yellow band centred at 1. In the legend, the $\chi^{2}$ per degree of freedom is given to quantify the goodness of fit of the model to the data.}
\label{fig:Z1J_dpt}
\end{figure}

For $\PZ$+ $\geq 1$ jet events, Fig.~\ref{fig:Z1J_dphi} (\ref{fig:Z1J_dpt}) shows the differential cross section measurements (left) and the area-normalized distributions (right) as a function of $\Delta \phi (\PZ, j_1)$ ($\Delta_\text{rel}\pt(\PZ,j_1)$), respectively. Different MC event generators (except for the MG5\_aMC + {\PYTHIA}8 with the DPS-specific tune CDPSTP8S1-WJ) describe, within the uncertainties, the overall differential cross section as a function of $\Delta\phi$ and $\Delta_\text{rel}\pt$, apart from a few discrepancies in some specific regions of these observables. MG5\_aMC + {\PYTHIA}8 generator prediction with the DPS-specific tune CDPSTP8S1-WJ overestimates the cross section up to 10--20\%, but correctly describe the shapes of the $\Delta\phi$ and $\Delta_\text{rel}\pt$ distributions. The MG5\_aMC + {\PYTHIA}8 prediction (with CP5 tune) overestimates (up to 20\%) the measurement in the lower-$\Delta_\text{rel}\pt$, where SPS is expected to be dominant. The prediction of  MG5\_aMC + {\HERWIG}7 describes, within the uncertainties, the shape of the $\Delta_\text{rel}\pt$ distribution, but deviates up to 15--20\% in the lower-$\Delta\phi$ region where DPS is expected to contribute more. 

The \zjets calculation of MG5\_aMC + {\PYTHIA}8 without MPI does not describe the measurement and is lower than the measurement by up to 50\% in both the lower $\Delta\phi$ and higher $\Delta_\text{rel}\pt$ regions, where the MPI contribution is expected to be the largest.

\begin{figure}[htbp]
\centering
\includegraphics[width=0.48\textwidth]{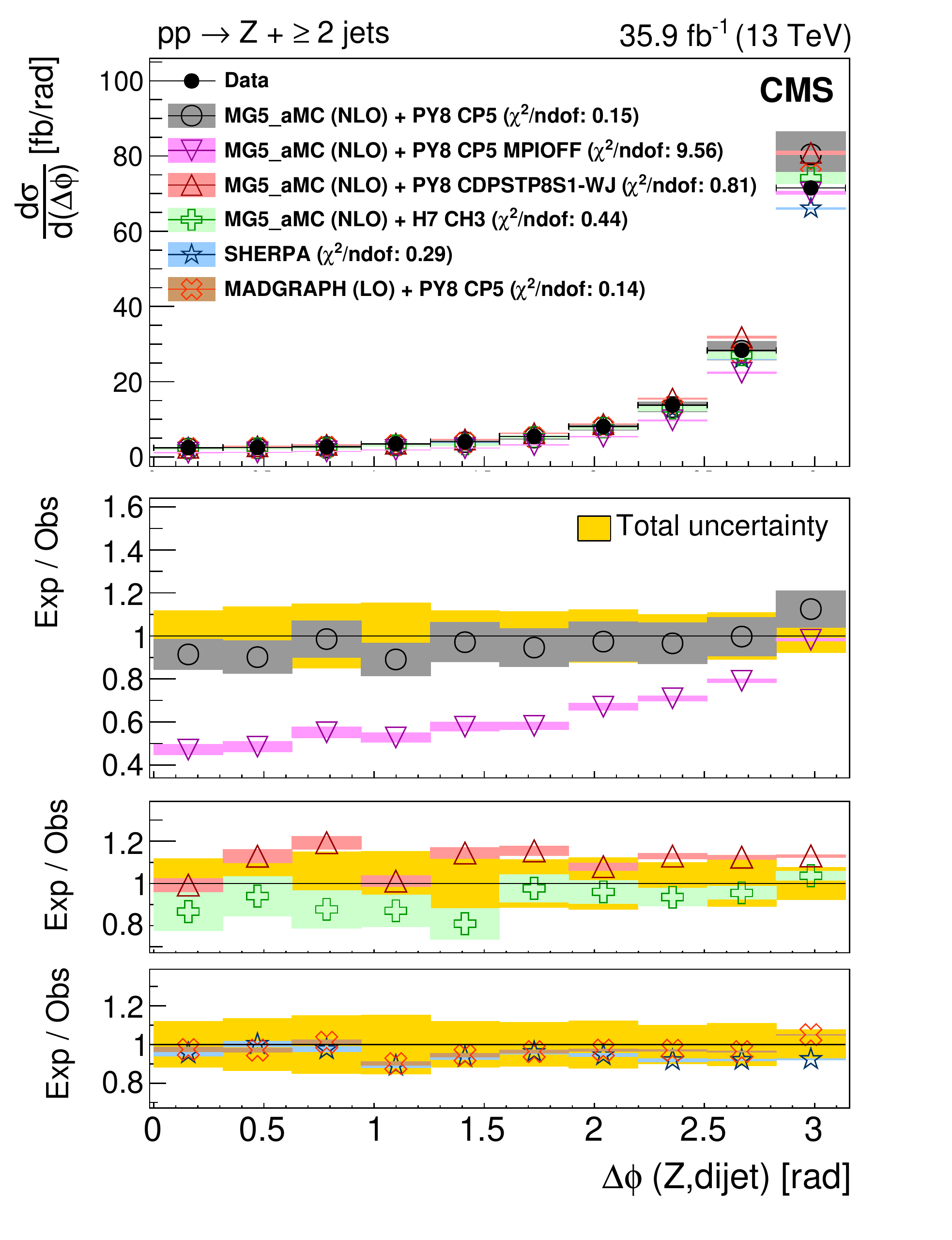}\label{fig:1dataMC_dphi_Zdijet_Z2J_CN_13TeV.pdf}
\includegraphics[width=0.48\textwidth]{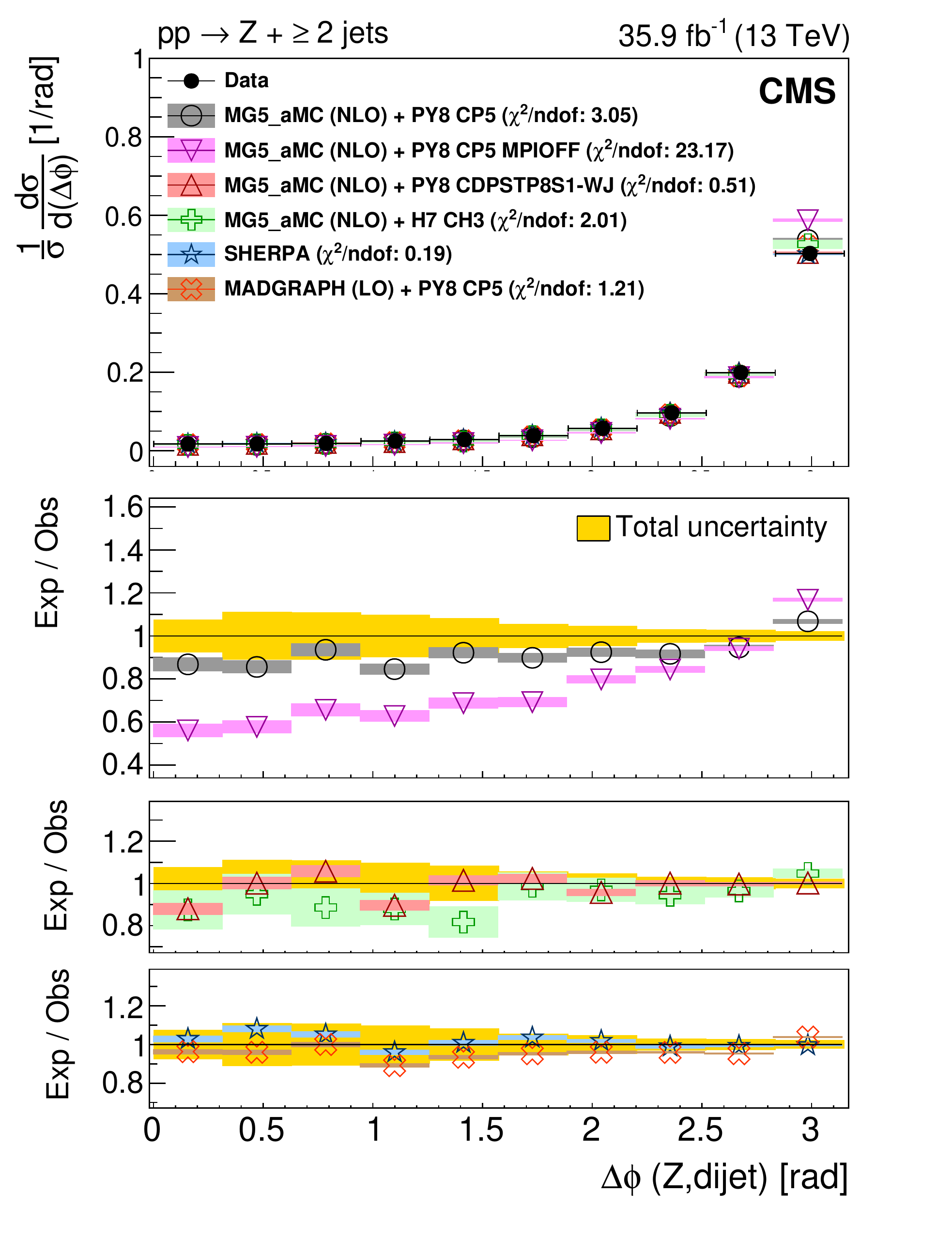}\label{fig:0dataMC_dphi_Zdijet_Z2J_SC_13TeV.pdf}
\caption{Differential cross sections (left) and area-normalized distributions (right) as functions of $\Delta\phi$ between the \PZ boson and the dijet for $\PZ$+ $\geq$2 jet events. The uncertainties in the predictions are shown as coloured bands around the theoretical predictions including statistical, PDF, scale and tune uncertainties for the NLO MG5\_aMC + {\PYTHIA}8 (with CP5 tune) and the statistical uncertainty only for the LO MG5\_aMC + {\PYTHIA}8 (with CP5 tune), NLO MG5\_aMC + {\PYTHIA}8 (with CDPSTP8S1-WJ tune, CP5 tune with MPI-OFF), NLO MG5\_aMC + {\HERWIG}7 (with tune CH3), and \SHERPA predictions. In the top panel, the vertical bars on the data points represent statistical uncertainties, whereas in the bottom panels, the total uncertainty in data is indicated by the solid yellow band centred at 1. In the legend, the $\chi^{2}$ per degree of freedom is given to quantify the goodness of fit of the model to the data.}
\label{fig:Z2J_dphi}
\end{figure}

\begin{figure}[htbp]
\centering
\includegraphics[width=0.48\textwidth]{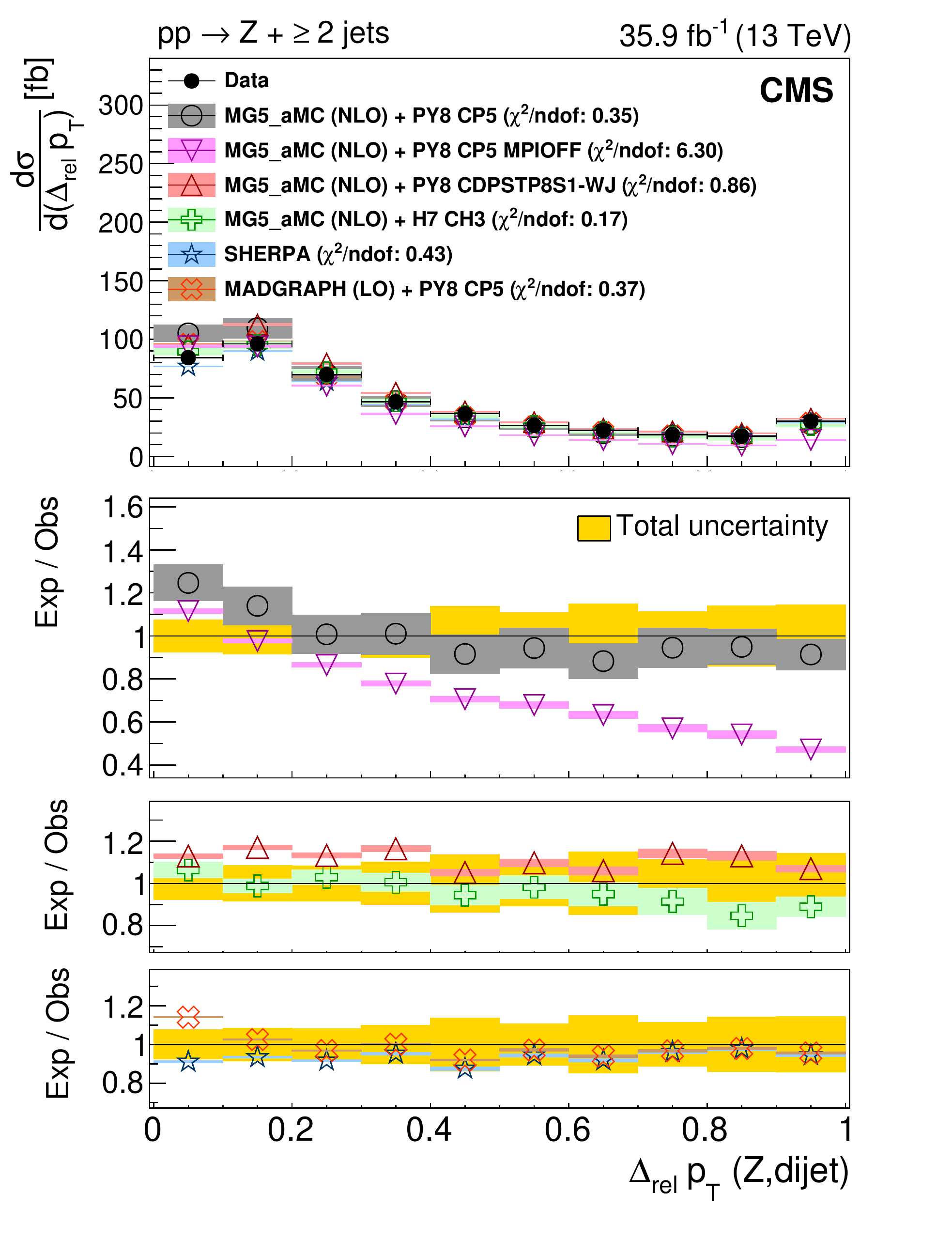}\label{fig:1dataMC_reldpt_Zdijet_Z2J_CN_13TeV.pdf}
\includegraphics[width=0.48\textwidth]{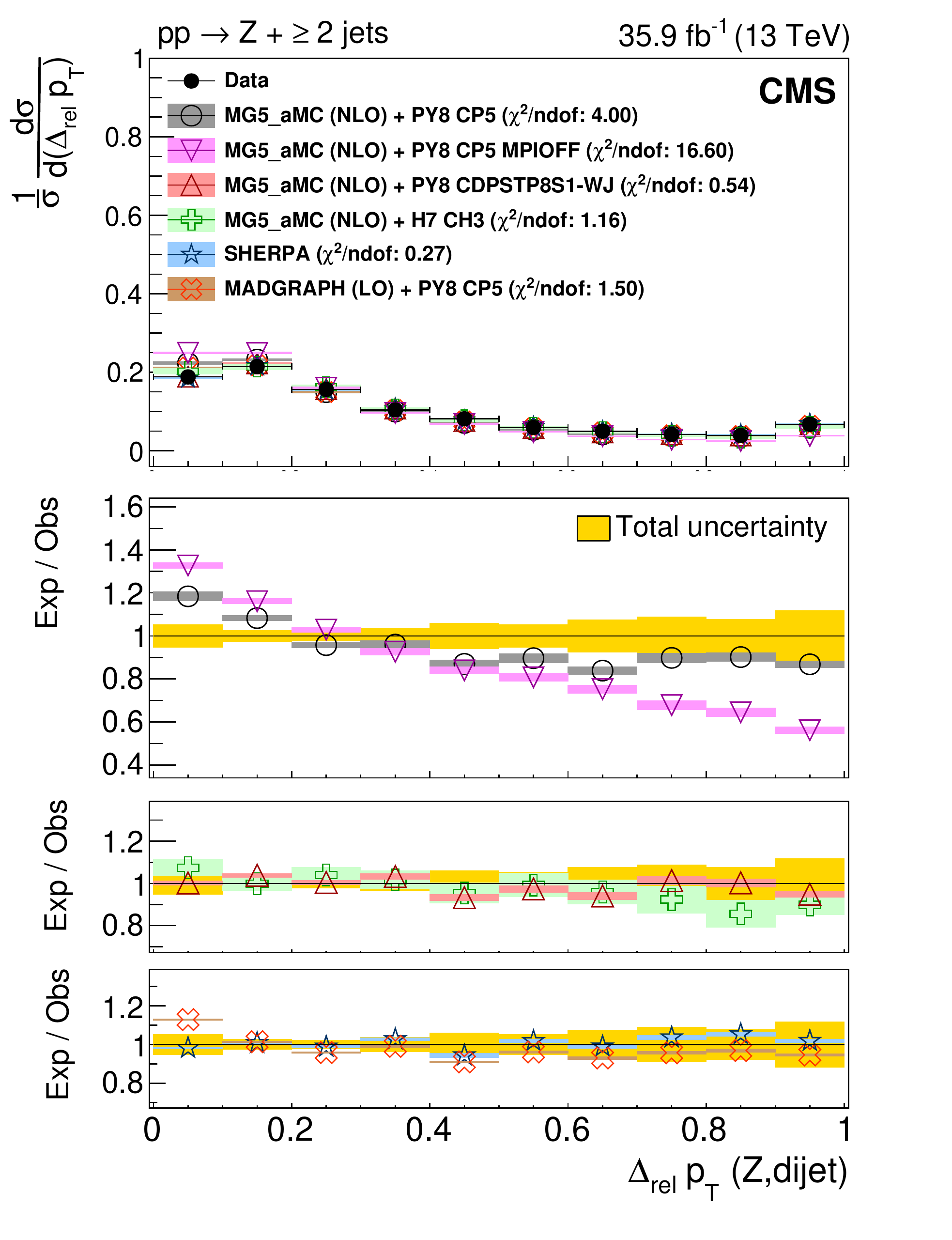}\label{fig:0dataMC_reldpt_Zdijet_Z2J_SC_13TeV.pdf}
\caption{Differential cross sections (left) and area-normalized distributions (right) as functions of the \pt imbalance between the \PZ boson and the dijet for $\PZ$+ $\geq$2 jet events. The uncertainties in the predictions are shown as coloured bands around the theoretical predictions including statistical, PDF, scale, and tune uncertainties for the NLO MG5\_aMC + {\PYTHIA}8 (with CP5 tune) and the statistical uncertainty only for the LO MG5\_aMC + {\PYTHIA}8 (with CP5 tune), NLO MG5\_aMC + {\PYTHIA}8 (with CDPSTP8S1-WJ tune, CP5 tune with MPI-OFF), NLO MG5\_aMC + {\HERWIG}7 (with tune CH3), and \SHERPA predictions. In the top panel, the vertical bars on the data points represent statistical uncertainties, whereas in the bottom panels, the total uncertainty in data is indicated by the solid yellow band centred at 1. In the legend, the $\chi^{2}$ per degree of freedom is given to quantify the goodness of fit of the model to the data.}
\label{fig:Z2J_dpt}
\end{figure}

{\tolerance=1200
For $\PZ$+ $\geq$2 jet events Fig.~\ref{fig:Z2J_dphi} (\ref{fig:Z2J_dpt}) shows the differential cross section measurements (left) and the area-normalized distributions (right) as a function of $\Delta \phi$(\PZ, dijet) ($\Delta_\text{rel}\pt(\PZ,\text{dijet})$), respectively. The differential cross section, as a function of $\Delta\phi$, is reasonably well described by the different predictions.
The predictions of MG5\_aMC + {\PYTHIA}8 (with the  CDPSTP8S1-WJ tune) and \SHERPA best describe the shape of the measured distributions, whereas the MG5\_aMC + {\PYTHIA}8 (with CP5 tune) prediction deviates by up to 15\% in the lower-$\Delta\phi$ region, where DPS production of \zjets is expected to contribute more.
The shape of the $\Delta_\text{rel}\pt$ distribution is best described by the predictions of MG5\_aMC + {\PYTHIA}8 (with the  CDPSTP8S1-WJ tune) and \SHERPA.
The MG5\_aMC + {\PYTHIA}8 (with CP5 tune) prediction overestimates, up to 20\%, in the lower-$\Delta_\text{rel}\pt$ region. 
\par}
 
\begin{figure}[htbp]
\centering
\includegraphics[width=0.48\textwidth]{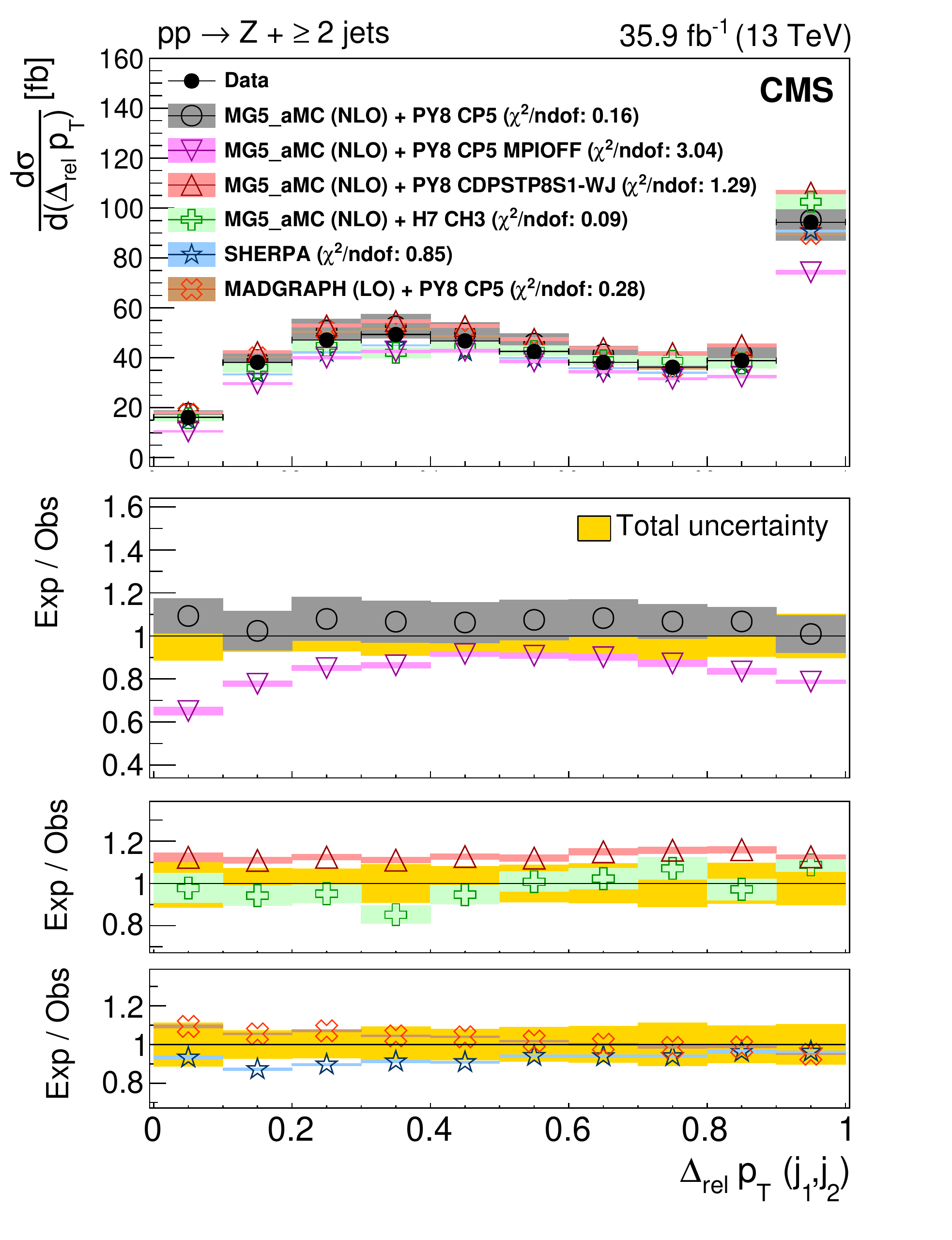}\label{fig:1dataMC_reldpt_j1j2_Z2J_CN_13TeV.pdf}
\includegraphics[width=0.48\textwidth]{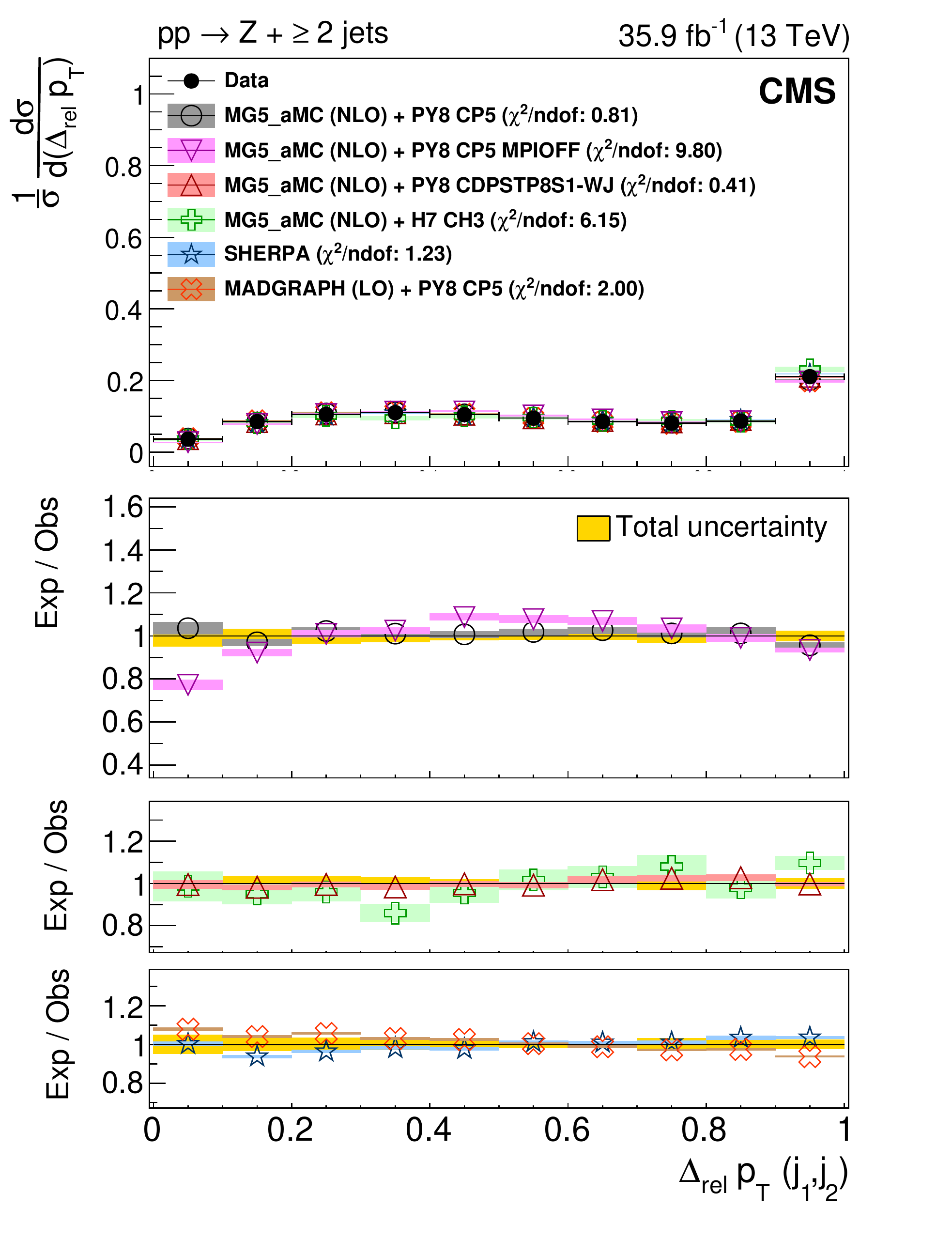}\label{fig:0dataMC_reldpt_j1j2_Z2J_SC_13TeV.pdf}
\caption{Differential cross sections (left) and area-normalized distributions (right) as functions of the \pt imbalance between leading and subleading jets for $\PZ$+ $\geq$2 jet events. The uncertainties in the predictions are shown as coloured bands around the theoretical predictions including statistical, PDF, scale, and tune uncertainties for the NLO MG5\_aMC + {\PYTHIA}8 (with CP5 tune) and the statistical uncertainty only for the LO MG5\_aMC + {\PYTHIA}8 (with CP5 tune), NLO MG5\_aMC + {\PYTHIA}8 (with CDPSTP8S1-WJ tune, CP5 tune with MPI-OFF), NLO MG5\_aMC + {\HERWIG}7 (with tune CH3), and \SHERPA predictions. In the top panel, the vertical bars on the data points represent statistical uncertainties, whereas in the bottom panels, the total uncertainty in data is indicated by the solid yellow band centred at 1. In the legend, the $\chi^{2}$ per degree of freedom is given to quantify the goodness of fit of the model to the data.}
\label{fig:j1j2_dpt}
\end{figure}

The differential cross section as a function of $\Delta_\text{rel}\pt$ between two jets, as shown in Fig.~\ref{fig:j1j2_dpt}, is well described by the different predictions except for MG5\_aMC + {\PYTHIA}8 (with the  CDPSTP8S1-WJ tune), which overestimates the differential cross section measurements up to 15\%. The shape of the $\Delta_\text{rel}\pt$ distribution is described well by the predictions presented, except some deviations shown by MG5\_aMC + {\HERWIG}7 mainly in the higher region of the distribution. If MPI is not included in the simulation, predictions underestimate the differential cross section and fail to describe the shape of all the observables for $\PZ$+ $\geq$2 jet events with deviations up to 50\%.

The model parameters for different MPI and hadronization models are mostly derived by fitting minimum bias and low-\pt ($\lesssim$3\GeV) MPI measurements. From the above data, it is clear that our measurements can test the accuracy of various predictions and the simulation of MPI. Most of the predictions follow a similar trend describing the differential cross sections and area-normalized distributions with a few exceptions.

The {\PYTHIA}8 CP5 tune (with MPI) describes the differential cross section measurements within the uncertainty, but deviate (up to 15--20\%) from the measurements in the lower-region of $\Delta_\text{rel}\pt(\PZ,j_1)$ and $\Delta_\text{rel}\pt(\PZ,\text{dijet})$, where SPS is expected to be dominant. In the case of area-normalized distributions, the prediction from the MG5\_aMC + {\PYTHIA}8 CP5 tune (with MPI) describes the shape of the $\Delta_\text{rel}\pt(j_1,j_2)$ distribution within the uncertainty, but overestimates in the lower-region of $\Delta_\text{rel}\pt(\PZ,j_1)$ and $\Delta_\text{rel}\pt(\PZ,\text{dijet})$, but underestimate otherwise. The LO calculations with MG5\_aMC + {\PYTHIA}8 provide a similar agreement as obtained by the NLO calculation. The MG5\_aMC + {\HERWIG}7 prediction describes the measurements within the uncertainty except some deviations in describing the shapes of the $\Delta \phi (\PZ, j_1)$ and $\Delta_\text{rel}\pt(j_1,j_2)$ distributions. The \SHERPA prediction is in reasonable agreement with the measurements within the uncertainty.

The predictions based on the DPS-specific tune CDPSTP8S1-WJ describe the shape of the distributions within the uncertainty. Since the parameters were derived by fitting only 7\TeV measurements, this suggests that the collision energy dependence of the MPI parameters is well modelled in this tune. 

\section{Summary} \label{sec_sum}
The CMS Collaboration has measured the differential cross sections for $\PZ$+ $\geq$1 jet and $\PZ$+ $\geq$2 jet events using proton-proton collision data at $\sqrt{s} $ = 13\TeV, corresponding to an integrated luminosity of 35.9\fbinv collected in the year 2016. The \PZ boson is reconstructed using the dimuon channel. This is the first measurement performed to explore observables sensitive to the presence of multi-parton interaction (MPI) using the \zjets process at 13\TeV. Within the fiducial region, the production cross sections of $\PZ$+ $\geq$1 jet and $\PZ$+ $\geq$2 jet events are measured to be $158.5\pm 0.3\stat\pm 7.0\syst\pm 1.2\thy\pm 4.0\lum$\unit{pb} and $44.8\pm 0.4\stat\pm 3.7\syst\pm 0.5\thy\pm 1.1\lum$\unit{pb}, respectively.
The measured integrated cross section in the fiducial region with jets is well described by the event generators \SHERPA, MG5\_aMC + {\PYTHIA}8, and MG5\_aMC + {\HERWIG}7 predictions.
The prediction obtained with MG5\_aMC + {\PYTHIA}8 with the double-parton scattering (DPS) specific tune CDPSTP8S1-WJ overestimates the measurements by 10--15\%, but correctly describes the shape of all the observables.
The prediction from MG5\_aMC + {\PYTHIA}8 with the CP5 tune, derived by fitting soft quantum chromodynamics (QCD) measurements, describes the differential cross section and area-normalized distributions. However, there are parts of the distributions that are not well described, such as single parton scattering that dominates lower regions of transverse momentum imbalance $\Delta_\text{rel}\pt$ distributions.
Predictions with other MPI models describe the measurements well (\SHERPA) or reasonably well (MG5\_aMC + {\HERWIG}7) except a few deviations in describing the shapes of the $\Delta \phi (\PZ, j_1)$ and $\Delta_\text{rel}\pt(j_1,j_2)$ distributions.
The measured distributions show a significant sensitivity to MPI. A proper simulation of MPI is essential to describe the shape of the measured distributions and hence these results are a useful input to further improve DPS-specific tunes and a global tune in combination with other soft QCD measurements.

\begin{acknowledgments}

  We congratulate our colleagues in the CERN accelerator departments for the excellent performance of the LHC and thank the technical and administrative staffs at CERN and at other CMS institutes for their contributions to the success of the CMS effort. In addition, we gratefully acknowledge the computing centres and personnel of the Worldwide LHC Computing Grid and other centres for delivering so effectively the computing infrastructure essential to our analyses. Finally, we acknowledge the enduring support for the construction and operation of the LHC, the CMS detector, and the supporting computing infrastructure provided by the following funding agencies: BMBWF and FWF (Austria); FNRS and FWO (Belgium); CNPq, CAPES, FAPERJ, FAPERGS, and FAPESP (Brazil); MES (Bulgaria); CERN; CAS, MoST, and NSFC (China); MINCIENCIAS (Colombia); MSES and CSF (Croatia); RIF (Cyprus); SENESCYT (Ecuador); MoER, ERC PUT and ERDF (Estonia); Academy of Finland, MEC, and HIP (Finland); CEA and CNRS/IN2P3 (France); BMBF, DFG, and HGF (Germany); GSRT (Greece); NKFIA (Hungary); DAE and DST (India); IPM (Iran); SFI (Ireland); INFN (Italy); MSIP and NRF (Republic of Korea); MES (Latvia); LAS (Lithuania); MOE and UM (Malaysia); BUAP, CINVESTAV, CONACYT, LNS, SEP, and UASLP-FAI (Mexico); MOS (Montenegro); MBIE (New Zealand); PAEC (Pakistan); MSHE and NSC (Poland); FCT (Portugal); JINR (Dubna); MON, RosAtom, RAS, RFBR, and NRC KI (Russia); MESTD (Serbia); SEIDI, CPAN, PCTI, and FEDER (Spain); MOSTR (Sri Lanka); Swiss Funding Agencies (Switzerland); MST (Taipei); ThEPCenter, IPST, STAR, and NSTDA (Thailand); TUBITAK and TAEK (Turkey); NASU (Ukraine); STFC (United Kingdom); DOE and NSF (USA).
   
  \hyphenation{Rachada-pisek} Individuals have received support from the Marie-Curie programme and the European Research Council and Horizon 2020 Grant, contract Nos.\ 675440, 724704, 752730, 765710 and 824093 (European Union); the Leventis Foundation; the Alfred P.\ Sloan Foundation; the Alexander von Humboldt Foundation; the Belgian Federal Science Policy Office; the Fonds pour la Formation \`a la Recherche dans l'Industrie et dans l'Agriculture (FRIA-Belgium); the Agentschap voor Innovatie door Wetenschap en Technologie (IWT-Belgium); the F.R.S.-FNRS and FWO (Belgium) under the ``Excellence of Science -- EOS" -- be.h project n.\ 30820817; the Beijing Municipal Science \& Technology Commission, No. Z191100007219010; the Ministry of Education, Youth and Sports (MEYS) of the Czech Republic; the Deutsche Forschungsgemeinschaft (DFG), under Germany's Excellence Strategy -- EXC 2121 ``Quantum Universe" -- 390833306, and under project number 400140256 - GRK2497; the Lend\"ulet (``Momentum") Programme and the J\'anos Bolyai Research Scholarship of the Hungarian Academy of Sciences, the New National Excellence Program \'UNKP, the NKFIA research grants 123842, 123959, 124845, 124850, 125105, 128713, 128786, and 129058 (Hungary); the Council of Science and Industrial Research, India; the Latvian Council of Science; the Ministry of Science and Higher Education and the National Science Center, contracts Opus 2014/15/B/ST2/03998 and 2015/19/B/ST2/02861 (Poland); the National Priorities Research Program by Qatar National Research Fund; the Ministry of Science and Higher Education, project no. 0723-2020-0041 (Russia); the Programa Estatal de Fomento de la Investigaci{\'o}n Cient{\'i}fica y T{\'e}cnica de Excelencia Mar\'{\i}a de Maeztu, grant MDM-2015-0509 and the Programa Severo Ochoa del Principado de Asturias; the Thalis and Aristeia programmes cofinanced by EU-ESF and the Greek NSRF; the Rachadapisek Sompot Fund for Postdoctoral Fellowship, Chulalongkorn University and the Chulalongkorn Academic into Its 2nd Century Project Advancement Project (Thailand); the Kavli Foundation; the Nvidia Corporation; the SuperMicro Corporation; the Welch Foundation, contract C-1845; and the Weston Havens Foundation (USA).
\end{acknowledgments}

\bibliography{auto_generated}
\cleardoublepage \appendix\section{The CMS Collaboration \label{app:collab}}\begin{sloppypar}\hyphenpenalty=5000\widowpenalty=500\clubpenalty=5000\vskip\cmsinstskip
\textbf{Yerevan Physics Institute, Yerevan, Armenia}\\*[0pt]
A.~Tumasyan
\vskip\cmsinstskip
\textbf{Institut f\"{u}r Hochenergiephysik, Wien, Austria}\\*[0pt]
W.~Adam, J.W.~Andrejkovic, T.~Bergauer, S.~Chatterjee, M.~Dragicevic, A.~Escalante~Del~Valle, R.~Fr\"{u}hwirth\cmsAuthorMark{1}, M.~Jeitler\cmsAuthorMark{1}, N.~Krammer, L.~Lechner, D.~Liko, I.~Mikulec, P.~Paulitsch, F.M.~Pitters, J.~Schieck\cmsAuthorMark{1}, R.~Sch\"{o}fbeck, M.~Spanring, S.~Templ, W.~Waltenberger, C.-E.~Wulz\cmsAuthorMark{1}
% ...
\vskip\cmsinstskip
% ...
1:  Also at TU Wien, Wien, Austria\\
% ...
\end{sloppypar}
\end{document}